# Mitigating Smishing: Challenges and Future Work

Cori Faklaris
*Department of Software and Information Systems, College of Computing and Informatics*
*University of North Carolina at Charlotte*
Charlotte, NC USA
cfaklari@charlotte.edu

*Abstract*— U.S. Federal Trade Commission data for 2020-2022 shows that 23% of all fraud reports include Text as the contact method, with reported losses in the same period rising 279%, to $326 million. These numbers, and estimates of other unreported losses, have spurred our lab and others in industry and academia to direct resources to understanding and mitigating smishing. In this workshop paper, we describe three principal challenges in smishing mitigation that are distinct from traditional phishing: (1) limitations of mobile device affordances, (2) the complexity of cellular network infrastructure, and (3) cognitive and contextual factors in mobile device usage. We describe ideas for mitigating smishing, including a high-level overview of our lab's work to create and test interface designs for text-message recipients.

## I. Introduction

Today, nearly all U.S. consumers use mobile phones—97%, according to Pew Research Center [1]—and scammers have followed them there. During the COVID-19 pandemic, text messages increased dramatically as the delivery method for cyber scams [2]. According to Federal Trade Commission statistics, 23% of all fraud reports for 2020-2022 identified Text as the contact method, exceeded only by Phone Call (26%) [3]. The Text contact method is popularly called "smishing" (aka "SMiShing," or "SMShing")—a contraction of email "phishing" and "SMS" (for Short Message Service). And, like phishing, it pays off for attackers. FTC data for 2022 shows that consumers reported losses of $326 million to text scams, an increase of 279% since 2020 (Table I). Smishing losses that include publicly unreported attacks and users of specific websites or apps may be much higher. In 2021, users of the payment app Zelle may have lost as much as $440 million, according to a report from U.S. Senator Elizabeth Warren [4].

These numbers have spurred our lab and others in industry and academia to direct resources to the smishing problem. We have been working toward creating and testing interface improvements to help mobile users to identify and avoid smishing text messages—which must of necessity be tied to back-end systems for better filtering or flagging of suspicious texts. These are in the tradition of Hong 2012's summary of the three main ways to mitigate phishing attacks: "make things invisible" (ex: deploy machine learning on the back end to classify and filter out likely phish), develop better user interfaces, and provide effective training [5]. In the process, we have uncovered several challenges that must be addressed for effective smishing mitigation. We present these ideas and challenges below as an invitation for discussion and collaboration on pursuing viable mitigations for smishing.

TABLE I. Data from fraud reports by Text contact method, Consumer Sentinel Network Data Book 2020-22, Federal Trade Commission [a]

| Year | # of Reports | % of all reports with Contact Method [b] | % with a dollar loss reported | Total $ Lost (in millions) | Median $ Loss |
|---|---|---|---|---|---|
| 2020 | 334,524 | 27% | 5% | $86 | $800 |
| 2021 | 377,840 | 21% | 4% | $131 | $900 |
| 2022 | 321,374 | 22% | 6% | $326 | $1,000 |
| Average | 344,579 | 23% | 5% | $181 | $900 |

[a.] Data from 2019 and earlier does not break out Text as a contact method.

[b.] Other contact methods include, in order of percentage: Phone calls, Email, Websites or Apps, Social Media, Online Ad or Pop-Up, Mail, and Other (TV or radio, print, fax, in person, consumer initiated contact, and other methods consumers write in or that cannot be otherwise categorized). A further 39% of fraud reports did not specify a contact method.

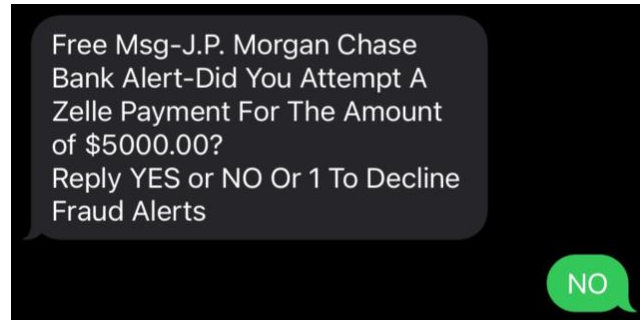

Fig. 1. Screenshot from a popular Zelle scam in which the attackers want to trick users not into falling for the first message, but for a faked response from the financial entity involved [10]. Security blogger Brian Krebs says that on any reply, even "No," the mobile user "will very soon after receive a phone call from a scammer pretending to be from the financial institution's fraud department. The caller's number will be spoofed so that it appears to be coming from the victim's bank."

## II. Background and Related Work

Smishing is a relatively newer form of phishing that uses fraudulent SMS text messages to deceive users into clicking on malicious links or providing personal information [6], [7], [8], [9]. The process commonly proceeds in 2-3 steps: (1) A fraudster sends a phishing link via Short Message Service (SMS) text to a phone. Then, the victim's click may (2) automatically start a money transfer, or (3) send them to a fake website form to provide sensitive info such as account passwords [10]. These messages often exploit users' emotions, such as fear, love, or greed, to induce them to take immediate action without verifying the source or the message [11]. Variations will trick recipients via a faked phone call from a fictional security team into giving up their account information (Fig. 1) [10].

Supported by the Center for Cybersecurity Analytics and Automation.

Banks, delivery companies, retailers, communication providers, and government agencies are commonly impersonated. For example, a smishing attack on customers of the U.S. Fifth Third Bank led them to enter their credentials on a bogus website, thinking the bank had requested this to unlock their accounts [12]. An even bigger attack tricked customers of Czech Post into downloading a malicious app to their phones [13]. Recently, attackers have exploited COVID-19 information confusion and the global shift to remote messaging to motivate users with bogus messages from contact tracing websites, insurance, or vaccine providers [1].

Though smishing and phishing are similar types of attacks, it is unclear the extent to which the existing research on phishing can be applied to smishing. One reason is the scarcity of academic research on who falls for smishing and why. Two 2023 exceptions are Faklaris, Lipford, and Tabassum [14], who found that about 18% of survey-takers fell for simulated smish text messages, and Rahman, Timko, Wali, and Neupane [11], who found that about 15% of people randomly served a smishing text message would respond to it. Notably, these are smaller percentages than the 52% of participants in Sheng, Holbrook, Kumaraguru, Cranor, and Downs's 2010 study who indicated that they would click on simulated phishing links [15]. The drop in gullible users may simply be the result of an increase in security awareness training in the past 15 years. It also may be that a rising volume of internet scams in general has increased users' ability to spot them (known as the "prevalence effect" [16]). However, the majority of mobile users simply now may default to judging any text message as a scam. Faklaris, Lipford, and Tabassum found that 76.5% of their study participants incorrectly judged their simulated real messages to be faked [14]. Similar to how the scam call epidemic has been undermining our telephony infrastructure [17], the sharp rise in smishing now may be eroding our ability to rely on text messaging for communication, too. More research is needed to determine whether these or other explanations are supported by data.

### III. Challenges in Mitigating Smishing

We have discussed the problem with knowledgeable industry specialists and reviewed available academic papers, white papers, reports, and blog posts. From this data, we have identified three major challenges in mitigating U.S. smishing that are distinct from traditional phishing and, in some cases, from how other countries or mobile ecosystems are approaching smishing. These are the limitations of mobile phone affordances for judging message credibility, the characteristics of U.S. cellular network infrastructure, and the different attitudes and environments in which mobile phones are known to be used.

#### A. Limitations of Mobile Device Affordances

Smishing attacks are more difficult for users to detect than email phishing attacks due to the technical limitations of the content form and the device interfaces [8]. Text messages themselves have fewer cues and indicators than emails, such as sender's address, subject line, or spelling errors [8]. And, several technical factors of mobile devices may constrict users' ability to assess and judge information, including: smaller screen sizes [18], [19]; limits on the ability to view sources and navigate among pages and apps (such as to view SSL certificates [20]), and a plethora of call-to-action interface elements [21], [22].

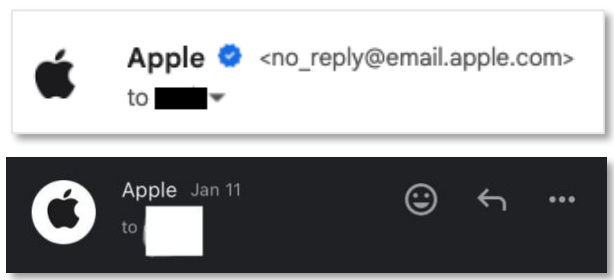

Fig. 2. A user of Gmail who views email on Google Chrome on a MacBook Pro (top image) will see much more information than on the Gmail app for iOS (bottom image). The desktop version shows the blue checkmark verification icon, while the mobile version does not, removing a key interface enhancement for guiding users' judgments of whether the email can be trusted. Also, the desktop version, but not the mobile version, will show the exact email address that the message was sent from. This is commonly taught as a way to check the legitimacy of an email message.

Even for responsive websites and apps, differences often exist between the desktop and the smaller-screen interfaces, which may mean that anti-phishing solutions developed for desktop turn out to not help mobile users. An example is the icon for verified senders in the Gmail header on desktop, which disappears in the Gmail mobile interface on iOS (Fig. 2).

Further, differences existing among the user interfaces of various mobile operating systems, with the market today split between Android and iOS [1], [23]. A notable example is the "blue bubble / green bubble" distinction among their text messaging apps, which, though seemingly minor, carries implications about the message parties' status and identity that have sparked strong emotions [24]. The lack of a standardized approach can lead to inconsistent mental models or varying levels of awareness among users across different mobile phone systems.

Mitigations developed for smishing in the mobile interface will need to account for and design around these device limitations. However, it also may be possible to make use of mobile device capabilities, such as haptic buzzes, or inferences drawn from mobile sensor data, such as whether someone is in a noisy environment or in motion and therefore multitasking, to help design smishing mitigations that leverage the phone environment and are tailored to the device context. Any interface designs or mobile notifications also will need data coming from a back-end system that is set up to detect whether incoming message are likely legitimate or are suspicious.

#### B. Complexity of Cellular Network Infrastructure

U.S. text messages are transmitted via radio signal to a network that might include several nodes or "hops" (Fig. 3) [25], [26], [27]. For the simplest types of messages, SMS, they are broken down into smallest bits of data and then converted to GSM 7-bit alphabet The GSM data is transmitted via radio signals to cell towers, then routed to a Short Message Service Center, which stores it until the receiver's phone connects to the network, then routes it to the receiver. The receiving phone then converts the GSM data back to readable text.

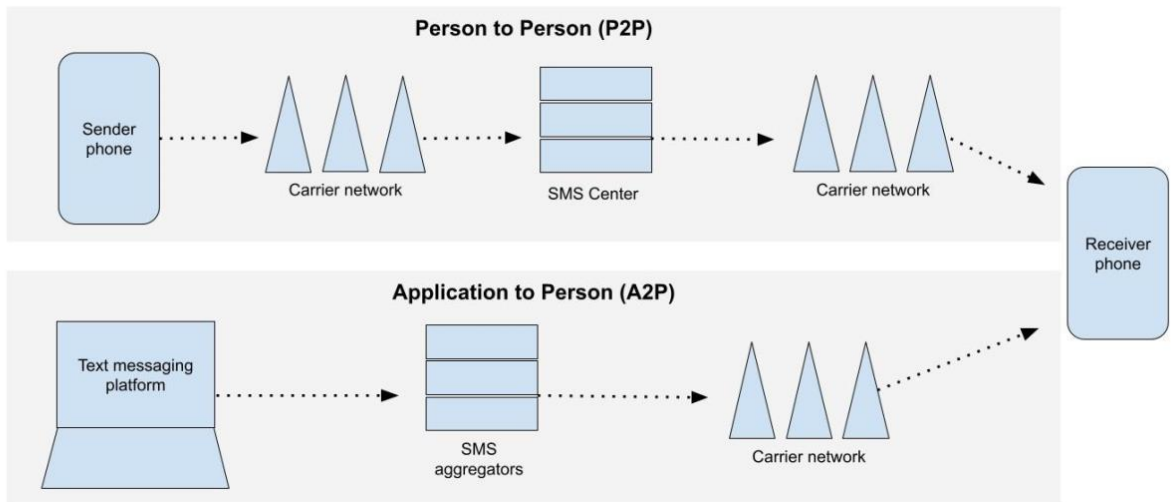

Fig. 3. Simplified view of how messages are transmitted from person to person (top image) and from application to person (bottom image) for mass messages. Each network checkpoint could be a place to insert a filter that helps spot and flag a suspicious text message, so that the receiver's phone can display a warning.

Several variables affect how messages route from sender to receiver: the message content, the mobile carriers for the recipient and for the sender, type of message, and whether special sender software is used. Beyond SMS, the type of message might be Multimedia Message Service (MMS) that allows a file attachment, Rich Communication Services (RCS) that adds more options for Android users, or Over-the-Top (OTT) for streaming video or other digital content [26]. (These differences in message type are partly why some message apps show "blue bubbles" and others are "green bubbles." [24]) Special sender software is used if the message is not person-to-person (P2P) but application-to-person (A2P), such as a commercial SMS gateway for mass communication of text messages.

Any back-end system designed to flag suspicious text messages will need to work regardless of mobile carrier, message type, or type of routing or gateway. While there are several checkpoints at which a smishing message might be detectable, not every message will go through the same sequence of checkpoints, nor are the same telecom companies or software providers involved consistently in the routing. It might be simpler to devise a U.S. system to note verified and/or legitimate senders, such as are already in place for email [28] and for voice calls [29], or to set up a naming scheme for legitimate senders' SMS short codes, such as the one instituted by India that adds a two-character prefix for such senders [30].

### C. Cognitive and Contextual Factors in Mobile Usage

A number of cognitive factors appear to be specific to mobile phone usage. These may make users more vulnerable to scams than if they were on desktop [22], [31]. First, text messages are more likely to be read and responded to than emails, as they are perceived as more personal and urgent [27]—leading marketers as well as scammers to send unsolicited texts to mobile numbers. Second, device owners are habituated by constant use to respond immediately to requests or react without reflection. Third, psychological factors that have been found to be associated with mobile users may leave them more open than desktop users to manipulation. These include increased cognitive overload and a greater willingness to self-disclose [32].

How these cognitive factors affect users may also vary according to context, such as whether a user is multitasking while using a smartphone [22], [31] or able to narrowly focus on a task but constrained by mobile's technical factors in their ability to make sense of information [32]. These factors may also vary by social group context, such as whether the phone is used in a multi-resident household [33], within a romantic relationship [34], or for work purposes [35].

Finally, our knowledge of cognitive and contextual factors may be constrained by the number of studies focused on populations who are close at hand: students, computer gig workers, and/or those who are receiving mobile stimuli in a lab experiment or through a survey or interview probe. More research is needed to look at specific, hard-to-survey populations and to test mitigations with them in the field. Such populations would include adults with poor eyesight, older users, less educated users, and people who are engaged in errands or weekend recreation while receiving text messages.

### IV. IDEAS FOR MITIGATION RESEARCH

The easiest smishing mitigation to design is one that will increase awareness of existing tools and safeguards. For example, we have found that many people are unaware of the 7726 (SPAM) number to which users can copy an unwanted message and forward, or of the ReportFraud.ftc.gov website for fraud reporting. One idea for increasing awareness would be to design and pilot-test an out-of-device marketing campaign to increase awareness among mobile phone users of these simple tools for reporting. Another idea is to design an in-device series of periodic notifications to remind mobile phone users of these reporting tools, such as a monthly text message or pop-up notification, or a notification that is triggered when someone uses the "Report Junk" or Delete function, or when they text a reply that is "Stop" or "No," to a sender who is not in their contact list, from their text messaging app of choice.

A similar mitigation idea, but more complex to program, would be to trigger an in-context warning or tip that cautions someone who is likely busy or multitasking to stop and think before engaging with an incoming text message. Such a system would take input from mobile sensors and shared data to calculate an activity inference (e.g. detecting via microphone decibels that someone is in a noisy place, or combining accelerometer and GPS data to determine that the phone and its user are walking around). Then, as a text message arrives, the system could overlay the text message with an interface signal such as a yellow "Caution" frame and icon. This is similar to how a red "Warning: Possible Phishing" box is appended to incoming emails from external senders in some enterprise email systems. It may have the effect of focusing the mobile user's attention, interrupting their instinct or habit to respond without reflection and causing them to think consciously about whether the incoming message is suspicious and how to deal with it.

Finally, our lab is exploring ideas for an in-app message indicator that can quickly and concisely communicate to the receiver how likely it is that the message is spam or worse. These ideas will make use of existing design patterns such as checkbox icons and knowledge such as color psychology of how to design for interface alerts that attract attention and are quickly understandable, without being overwhelming or annoying. These indicators will need to be carefully designed with attention to screen sizes, minimum icon sizes and type sizes that are recognizable regardless of visual acuity. To add accessibility for the blind and visually impaired and layer a different stimulus on top of the visual signal, these also could be designed to send a haptic signal such as a short and unique buzz when the user focuses on it.

The latter ideas, in particular, will rely also on a back-end system for filtering and for flagging suspicious text messages. This might be adaptable from the existing network affordances for the STIR-SHAKEN framework that supports Caller ID and helps flag robocalls [29]. It could also make use of similar standards and certificates to those used for Gmail sender verification [28]: DMARC (Domain-based Message Authentication, Reporting, and Conformance), which supports email systems' ability to recognize and separate legitimate emails from spoofed ones; BIMI (Brand Indicators for Message Identification), an open standard that supports use of brand logos to boost credibility; and VMC (Verified Mark Certificate), an authority-administered digital certificate that verifies logo ownership. However, this type of mitigation will require buy-in from key U.S. network stakeholders, chiefly mobile carriers, the SMS marketing industry, and the Federal Communication Commission, which has already taken initial steps to reduce robotexts and the misuse of unassigned phone numbers [36].

## V. CONCLUSION

In this workshop paper, we have summarized the urgency of research to understand and mitigate smishing. We described three major challenges in mitigating U.S. smishing that are distinct from traditional phishing: the limitations of mobile phone affordances for judging message credibility, the characteristics of U.S. cellular network infrastructure, and the cognitive and context factors that affect mobile usage. Finally, we briefly set forth ideas for mitigation research, some of which are easy to design and test without industry collaboration, and some of which will require the involvement of regulators, mobile carriers, and the marketing industry. We hope this work will spur discussion about next steps and possible future collaborations to address smishing. We believe that mitigating this problem is critical to preserve the usability of the U.S. mobile communication infrastructure.

ACKNOWLEDGMENT

Many thanks to Sarah Tabassum and Narges Zare for their comments and suggestions on this draft, and to Heather Richter Lipford and other collaborators for their input on the research.